\documentclass[sigconf,authorversion,nonacm]{acmart}
\usepackage{multirow}
\usepackage{multicol}
\usepackage[greek,english]{babel}
\usepackage[T1]{fontenc}

\AtBeginDocument{%
  \providecommand\BibTeX{{%
    Bib\TeX}}}

\setcopyright{acmcopyright}
\copyrightyear{2023}
\acmYear{2023}
\acmDOI{XXXXXXX.XXXXXXX}

\acmConference[JCDL '23]{Joint Conference on Digital Libraries (JCDL)}{June 26 - 30, 2023}{Santa Fe, New Mexico, USA}
\acmPrice{15.00}
\acmISBN{978-1-4503-XXXX-X/18/06}

\begin{document}

\title{Yes but.. Can ChatGPT Identify Entities in Historical Documents?}

\author{Carlos-Emiliano González-Gallardo}
\email{carlos.gonzalez_gallardo@univ-lr.fr}
\orcid{0000-0002-0787-2990}
\affiliation{%
  \institution{University of La Rochelle, L3i}
  \streetaddress{Avenue Michel Crépeau}
  \city{La Rochelle}
  \country{France}
  \postcode{F-17000}
}
\author{Emanuela Boros}
\authornote{This work was done while at University of La Rochelle, in La Rochelle, France.}
\email{emanuela.boros@epfl.ch}
\orcid{0000-0001-6299-9452}
\affiliation{%
  \institution{Digital Humanities Laboratory, EPFL}
  \city{Vaud}
  \country{Switzerland}
}

 \author{Nancy Girdhar}
 \email{nancy.girdhar@univ-lr.fr}
 \affiliation{%
  \institution{University of La Rochelle, L3i}
  \streetaddress{Avenue Michel Crépeau}
  \city{La Rochelle}
  \country{France}
  \postcode{F-17000}
}
\author{Ahmed Hamdi}
\email{ahmed.hamdi@univ-lr.fr}
\orcid{0000-0002-8964-2135}
\affiliation{%
  \institution{University of La Rochelle, L3i}
  \streetaddress{Avenue Michel Crépeau}
  \city{La Rochelle}
  \country{France}
  \postcode{F-17000}
}

\author{Jose G.	Moreno}
\email{jose.moreno@irit.fr}
\orcid{0000-0002-8852-5797}
\affiliation{%
  \institution{University of Toulouse, IRIT}
  \city{Toulouse}
  \country{France}
}
\author{Antoine Doucet}
\email{antoine.doucet@univ-lr.fr}
\orcid{0000-0001-6160-3356}
\affiliation{%
  \institution{University of La Rochelle, L3i}
  \streetaddress{Avenue Michel Crépeau}
  \city{La Rochelle}
  \country{France}
  \postcode{F-17000}
}

\renewcommand{\shortauthors}{González-Gallardo et al.}

\begin{abstract}
Large language models (LLMs) have been leveraged for several years now, obtaining state-of-the-art performance in recognizing entities from modern documents. For the last few months, the conversational agent ChatGPT has ``prompted'' a lot of interest in the scientific community and public due to its capacity of generating plausible-sounding answers. In this paper, we explore this ability by probing it in the named entity recognition and classification (NERC) task in primary sources (e.g., historical newspapers and classical commentaries) in a zero-shot manner and by comparing it with state-of-the-art LM-based systems. Our findings indicate several shortcomings in identifying entities in historical text that range from the consistency of entity annotation guidelines, entity complexity, and code-switching, to the specificity of prompting. Moreover, as expected, the inaccessibility of historical archives to the public (and thus on the Internet) also impacts its performance.

\end{abstract}

\begin{teaserfigure}
\centering
  \includegraphics[width=.9\textwidth]{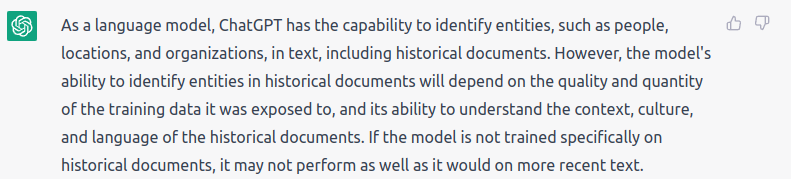}
  \caption{Asking ChatGPT to identify named entities in a historical document.}
  \Description{ChatGPT answer to a historical document.}
  \label{fig:teaser}
\end{teaserfigure}


\maketitle

\section{Introduction}

Since OpenAI\footnote{\url{https://openai.com/}} released ChatGPT at the thirty-sixth conference on neural information processing systems (NeurIPS) in November 2022\footnote{\url{https://nips.cc/}}, its ability to provide human-like and plausible-sounding answers caused the model to become extremely popular beyond the research community, gaining more than 1 million users in less than one week. ChatGPT is a conversational agent based on GPT-3.5 (generative pretrained transformer), a large language model (LLM) with more than 175 billion parameters \cite{ouyang2022training}. Given its widespread popularity and accessibility, the question of how this highly mediatized model performs on different natural language processing (NLP) tasks arose already in several fields \cite{biswas2023chatgpt,pavlik2023collaborating}.


\begin{table*}[ht]
\begin{center}
\caption{Dataset prompts used to collect the predictions.}
\addtolength{\tabcolsep}{2.6pt}

\begin{tabular}{ p{5.5cm} p{5.3cm} p{5.5cm} }
\hline
    \multicolumn{1}{c}{{NewsEye}} & \multicolumn{1}{c}{\texttt{hipe-2020}} & \multicolumn{1}{c}{\texttt{ajmc}} \\ \hline

What are the locations (LOC), persons (PER),
organizations (ORG) and human productions (HumanProd)
present in the following historical text? 
\textbf{\textit{\{SENTENCE\}}}
Respond, for each word, using IOB or BIO format separated
by tab. If a word has no entity, add O.
&
What are the locations (loc), persons (pers),
organizations (org), products (prod) and time periods
(time) present in the following historical text?
\textbf{\textit{\{SENTENCE\}}} 
Respond, for each word, using IOB or BIO format separated
by tab. If a word has no entity, add O. 
&
What are the locations (loc), persons (pers), time
periods (date), human works (work), physical objects
(object) and specific portion of works (scope) present
in the following historical text?
\textbf{\textit{\{SENTENCE\}}}
Respond, for each word, using IOB or BIO
format separated by tab. If a word has no entity, add O. \\ \hline
\end{tabular}
\label{table:chatgpt_prompts}
\end{center}
\end{table*}

LLMs have been leveraged for several years now, obtaining state-of-the-art performance in the majority of NLP tasks, by generally being fine-tuned on downstream tasks such as named entity recognition and classification (NERC) and less in zero-shot settings \cite{li2020survey}. Thus, for NERC, but also as a general focus, efforts are dedicated to how to effectively transfer knowledge for domain adaptation by developing cross-domain robust systems and exploring zero-shot, or few-shot learning to address domain and annotation consistency and mismatch in cross-domain settings \cite{ehrmann2020extended,ehrmann2022extended}. Simultaneously, in historical documents (e.g., historical newspapers and broadcasts), NERC faces new challenges apart from domain heterogeneity such as input noisiness, language dynamics and lack of resources \cite{ehrmann2021named}. 

Processes such as optical character recognition (OCR) or optical layout recognition (OLR) affect consistently the transcriptions and thus, this propagated noise influences the precision of NERC systems \cite{schweter-baiter-2019-towards,gonzalez2023injecting,borocs2020alleviating,najem2022page,schweter2022hmbert,boros2022knowledge}. Even though the latest developments in deep learning by fine-tuning and pre-training historical LMs brought state-of-the-art results in NERC in historical documents \cite{fonteyn2022adapting,aprosio2022bertoldo}, time and domain shifts and resource scarcity remain crucial challenges in learning or reusing appropriate knowledge for NERC. Thus, as expected, LLMs and systems in which they are embedded such as ChatGPT were not explicitly trained for information extraction tasks \cite{qin2023chatgpt} (e.g., named entity recognition and classification, relation extraction), and moreover, as seen in Figure \ref{fig:teaser}, not specifically with a focus in historical documents.


In this short preliminary work, we conduct an exploratory case study to investigate the potential of ChatGPT, which was trained on a massive amount of Internet data (e.g., Common Crawl, WebText2, Wikipedia) \cite{brown2020language} and prompt datasets for reinforcement learning from human feedback (RLHF) \cite{ouyang2022training}. Due to this increase in scale in comparison with previous generative language models such as GPT-3 and GPT-2 \cite{brown2020language,floridi2020gpt}, the behaviour of the model drastically changed, being considerably more able to perform tasks it was not explicitly trained on than previous models \cite{perez2021true}. We would expect ChatGPT to be able to detect entities in historical documents to a certain degree, considering the aforementioned challenges, and therefore, we conduct this study by experimenting with zero-shot NERC and comparing its performance against state-of-the-art systems.


\section{Datasets}

We selected three historical document datasets covering circa 200 years, they are composed of classical commentaries and historical newspapers, provided by digital libraries during different international research projects, i.e., NewsEye\footnote{\url{https://www.newseye.eu/}} and \textit{impresso}\footnote{\url{https://impresso-project.ch/}}.


The {NewsEye} dataset \cite{hamdi2021multilingual} was collected through the national libraries of France (BnF), Austria (ONB) and Finland (NLF)\footnote{Bnf: \url{https://bnf.fr}; ONB: \url{https://onb.ac.at}; NLF: \url{https://kansalliskirjasto.fi}}. It comprises four corpora (French, German, Finnish, and Swedish), being the {French} one composed of texts from digitized archives of nine newspapers (i.e., \textit{L'Oeuvre}, \textit{La Fronde}, \textit{La Presse}, \textit{Le Matin}, \textit{Marie-Claire},\textit{Ce soir}, \textit{Marianne}, \textit{Paris Soir} and \textit{Regards}) from 1854 to 1946.

\begin{table}
\begin{center}
\caption{Statistical description of corpora (PERS = person, LOC = location, ORG = organization, HumanProd = human work/production, TIME = date/interval, SCOPE = specific portion of work).}    
\setlength{\tabcolsep}{1.1pt}
\begin{tabular}{ c c c c c c c c }
\hline
\bf Tokens & \bf Entities & \bf PERS & \bf LOC & \bf ORG & \bf HumanProd & \bf TIME & \bf SCOPE\\
\hline

\hline
\multicolumn{8}{c}{{NewsEye}} \\ \hline
 30,458 & 1,298 & 463 & 597 & 217 & 21 & -- & --\\
\hline
\multicolumn{8}{c}{\texttt{hipe-2020}} \\   \hline
48,854& 1,600 & 502 & 854 & 130 & 61 & 53 & -- \\
\hline   
\multicolumn{8}{c}{\texttt{ajmc}} \\   \hline
5,390 &360 & 139 & 9 & -- & 80 &3 & 129\\
\hline
\end{tabular}
\label{ner_dataset}
\end{center}
\end{table}

The {\texttt{hipe-2020}} dataset \cite{ehrmann2020language} is composed of Swiss, Luxembourgish and American newspaper articles in French, German and English comprising the 19th and 20th centuries. It has been collected mainly through the National Library of Switzerland (BN), the National Library of Luxembourg (BnL), the Media Center and State Archives of Valais and the Swiss Economic Archives (SWA)\footnote{BN: \url{https://www.nb.admin.ch}; BnL: \url{https://bnl.public.lu}; SWA: \url{https://wirtschaftsarchiv.ub.unibas.ch}} as part of the \textit{impresso} project.


The {\texttt{ajmc}} dataset \cite{matteo2021optical} is composed of classical commentaries from the Ajax Multi-Commentary project that includes digitized 19th century commentaries published in French, German, and English. These commentaries provide in-depth analysis and explanation of Sophocles' Ajax Greek tragedy.

These datasets have been annotated with universal (i.e., person, location, organization) and domain-specific (i.e., bibliographic references to primary and secondary literature) entity types and subtypes for the NERC task and split into train, development and test partitions. During this preliminary work, only French test partitions have been taken into consideration. Table \ref{ner_dataset} displays the information regarding the number and type of entities found in the specified datasets. 

\section{Methodology}

We followed a straightforward zero-shot approach to retrieve named entities from ChatGPT via the official web interface\footnote{\url{https://chat.openai.com}} between January 11th and February 7th, 2023. An upgrade was released on January 30th to improve factuality and mathematical capabilities of the model\footnote{\url{https://help.openai.com/en/articles/6825453-chatgpt-release-notes}}, however, we did not perceive any difference with regard to the ability of the model to detect entities. 

\texttt{hipe-2020} and \texttt{ajmc} were provided with fine-grained and nested entities, but in order to simplify the complexity of the prompts for this study, we only consider the coarse-grained entities. We defined the three prompts presented in Table \ref{table:chatgpt_prompts} depending on the entity type differences between datasets and to respect their corresponding labels casing. Even if the IOB/BIO\footnote{\url{https://en.wikipedia.org/wiki/Inside\%E2\%80\%93outside\%E2\%80\%93beginning\_(tagging)}} format is explicitly demanded for each word, tokenization by ChatGPT was inconsistent with IOB tokenized dataset files, thus, for evaluation purposes, an 
alignment and verification process was required to ensure evaluation consistency.

%


\section{Results}

Table \ref{tbl:chatgptresults} presents the performance of ChatGPT over coarse-grained (high-level entity types) NERC in terms of strict and fuzzy micro-level precision (P), recall (R) and F-measure (F1) evaluated with CLEF-HIPE-2020-scorer\footnote{\url{https://github.com/hipe-eval/HIPE-scorer}}. For positioning and comparison, we also present the performance of two LM-based state-of-the-art NERC systems.

\begin{table}[ht]
\begin{center}
\caption{Comparative results using the three datasets (micro).}
\addtolength{\tabcolsep}{-3pt}\begin{tabular}{ l c c c |c c c |c c c }
\hline
    &  \multicolumn{3}{c|}{{NewsEye}} & \multicolumn{3}{c|} {\texttt{hipe-2020}} & \multicolumn{3}{c}{\texttt{ajmc}} \\ \hline & P & R & F1 & P & R & F1 & P & R & F1 \\ \hline
 & \multicolumn{9}{c}{strict} \\ \hline
\textit{Stacked NERC} &  \bf 75.0 &  70.6 & \bf 72.7 & -- & -- & -- & --& --& -- \\ 
\textit{Temporal NERC} & -- & -- & -- & \bf 76.5 & \bf 76.5 & \bf 76.5 & \bf 84.8 & \bf 83.9 & \bf 84.4 \\ 
 ChatGPT  &  70.9  & \bf 72.3  & 71.6  & 32.5  & 50.0 & 39.4 & 21.8 & 26.1 & 23.8  \\\hline
 & \multicolumn{9}{c}{fuzzy} \\ \hline
\textit{Stacked NERC} & \bf 85.4 & \bf 80.5	& \bf 82.9 & -- & -- & --& --& --& -- \\ 
\textit{Temporal NERC}  & -- & -- & - & \bf 86.7 & \bf 86. 7& \bf 86.7 & \bf 90.2 & \bf 89.2 & \bf 89.7 \\
 ChatGPT  & 77.8  & 79.4  & 78.6  & 49.0  & 75.4 & 59.4 & 25.5 & 30.6 & 27.8 \\ \hline
\end{tabular}
\label{tbl:chatgptresults}
\end{center}
\end{table}

\paragraph{Stacked NERC} This model is based on the pre-trained model BERT proposed by \cite{devlin2018bert} with a stack of $2$ Transformer blocks on top, finalized with a conditional random field (CRF) prediction layer. \textit{Stacked NERC} proved increased performance in historical documents, while did not degrade the performance over modern data \cite{boros2020robust,borocs2020alleviating}. The same architecture was utilized as a baseline in the description of the NewsEye dataset \cite{hamdi2021multilingual}.

\paragraph{Temporal NERC} This model relies on \textit{Stacked NERC}, and it includes a data-wise improvement by exploiting temporal knowledge graphs for generating additional contextual time information and a model-wise improvement that incorporates this information with mean-pooled context jokers \cite{gonzalez2023injecting}. \textit{Temporal NERC} proved the importance of temporality for historical newspapers and classical commentaries, depending on the time intervals and the digitization error rate.

From Table \ref{tbl:chatgptresults} it is clear that the capacity of ChatGPT of  identifying named entities is really dependent on the dataset and the type of entities. Noticeable drops in performance can be observed for \texttt{ajmc} with over 71\% decrease for strict and over 69\% for fuzzy metrics. For \texttt{hipe-2020}, the performance has decreased less drastically, with over 48\% for strict and 31.48\% for fuzzy. With respect to NewsEye, the score values are marginally similar, with a drop of close to 1.5\% for strict and slightly over 5\% for fuzzy. While the results are generally balanced, we also observe a higher recall in the case of \texttt{hipe-2020} which could indicate that the complexity of the entity annotation causes ChatGPT to be able to detect them but not correctly classify them.

%
%


\section{Limitations}
We refer to limitations as the weaknesses of which ChatGPT gives evidence in the process of recognizing entities in a historical text, and less regarding limitations of this work which are not in our control, such as the inability of exploring the insights of ChatGPT and ensuring reproducible results if the same method is applied\footnote{All predictions are available at \url{https://anonymous.4open.science/r/NERC_ChatGPT-F687/}.}. We, thus, study the impact of the quality of the datasets and their annotations with regard to the ChatGPT responses.

\subsection{Named Entity Definition} 

Named entity annotation follows well-defined guidelines to describe the nature and boundaries of universal and domain-specific entity types, yet it is necessary to rely on the linguistic intuition and awareness of the annotator \cite{hamdi_2021_guidelines_newseye,ehrmann_2020_guidelines_hipe,romanello_2022_guidelines_ajmc}. 
While the definitions of universal entity types are stable (e.g., person, location and organization), domain-specific entity types vary among guidelines.

\texttt{hipe-2020} \cite{ehrmann_2020_guidelines_hipe} and NewsEye \cite{hamdi_2021_guidelines_newseye} guidelines define a ``human production'' entity as anything that is broadcast in the press, on radio or television such as newspapers, magazines, broadcasts, sales catalogues, etc. (e.g., ``Die Zeit'', ``Le Figaro'', ``Le sept à huit'', ``La ferme célébrités'') and exclude media products such as films, TV series, etc., and political, philosophical, religious/sectarian doctrines (e.g., ``Der Sozialismus'', ``Theravada Buddhismus'', ``Le socialism'', ``le bouddhisme theravâda'').



Similarly, a ``work'' entity is described by the \texttt{ajmc} guidelines \cite{romanello_2022_guidelines_ajmc} as an entity denoting a human creation, be it intellectual or artistic, that can be referred to by its title; ``A work is a distinct intellectual or artistic creation'' (FRBR\footnote{\url{https://www.oclc.org/research/activities/frbr.html}} guidelines) including literary works, religious works, editions of papyrological and epigraphical sources (e.g. ``IG2'', ``P.Oxy 1.119''), and journals. 

All descriptions overlap, but they undoubtedly lead to different annotations specific to each dataset, that could lead to slight confusions in predictions. For example, in \textit{\textsf{Le sens est donc : « Ai-je parlé trop peu clairement ? » Cf. Antigone, 405 : “Ap’ Eväênda xxi cœpñ XéYw ; Eschyle, Agamemnon, 269 : ”\textgreek{H xoρῶς λέγω} ;}.},  an example from \texttt{ajmc}, where the word \textit{Antigone} refers to a human creation, the entity is not detected by ChatGPT. However, since we are exploring the model in a zero-shot manner, and while we agree that consistency of the annotations is a major concern because of the language ambiguity, we can only assume that this misclassification comes more from the lack of specificity in the prompt than the entity definition.

\subsection{Entity Complexity} 

For NewsEye guidelines, a named entity is a real-world object denoting a unique individual with a proper name. Since, historically, a person's name played an influential role in reflecting key attributes of their job or life, the majority of entities have also been annotated including the person's job title. In the following example:
\textit{\textsf{Becthoven. — Par le Quatuor de la fondation Beethoven : MM.A, Géioso1er violon ; A. Tracol ; 2e violon ;P. Monteux. aito, F. Schnéklud, violon-celle ; César Geloso, pianiste.}} César Geloso, the pianist, is considered a ``person'' entity, however, ChatGPT is unable to detect beyond the mention of the first and last names. If jobs are needed to be detected inside entities, we assume that more information should be added in the prompt.


If the first or last name are in the text, ChatGPT seems to give them priority because it is able to identify the presence of an entity in the text, even though no names are mentioned (e.g., \textit{\textsf{garçon, infirmière des Hôpitaux}}). Thus, in \textit{\textsf{Ouvrier d'art et artiste peintre cherche petits travaux 
restauration de tableaux, décoration d'intérieur ou 
réparations meu¬bles de style, anciens ou modernes, 
laques, vernis Martin et au tampon. Christophe, 56, 
rue de la Montagne-Sainte-Geneviève, 5e 43 ans.}} the art worker and painter that looks for small jobs is correctly spotted. As well, in \textit{\textsf{
Un homme à grosses	moustaches, pensée, sourit soudain 
imperceptible¬	ment. Mais La jeune femme du premier
rang au manteau de vison, jusqu’alors muette.}}, the man with a big moustache was also detected, along with the young woman of the first rank.

Letter casing also seems to be challenging. While it is common for the names of organizations and headlines of newspaper articles to be all capitals (e.g., LE SPORT, NOUVELLES BREVES), ChatGPT identified them all as organizations. In the case of demonyms, locations and persons, they were systematically mixed up (e.g., \textit{\textsf{Carcassonnais, Russe, Mexicains, ``Italien, 17 3/4'', ``Japon 1899, 73'', ``Portugais 3 \%, 2 1/4'', ``Russe 1906''}}). It is unclear, however, if the confusion comes from the fact that they start with capital letters or due to context, as this limitation is commonly found in recent state-of-the-art NERC models.

NewsEye guidelines consider addresses such as \textit{\textsf{130, rue de la Courselle}} and \textit{\textsf{ 56, rue de la Montagne-Sainte-Geneviève, 5e}} as locations, however, ChatGPT does not seem to capture this fine level of granularity. Since the word ``location'' defines space in a more semantically generic manner than a specific place or position, an address is referencing the particulars of a place, which, if unspecified in the prompt, ChatGPT is unable to correctly identify.

\subsection{Digitization Errors}


Quantitatively, ChatGPT identified only 7\% of named entities with OCR errors in the \texttt{ajmc} dataset, while \textit{Temporal NERC} correctly identified around 40\% of noisy entities. \textit{Temporal NERC} recognized named entities with up to 70\% of their characters sustained an OCR error (i.e., deletion, insertion, substitution), however, character errors should not exceed 20\% on entities to be recognized by ChatGPT. More specifically, ChatGPT showed an inability to recognize named entities with segmentation errors. For instance, in \textit{\textsf{13659 - 4360.Hxépra. … . elloug. Ulysse paraît faire allusion à l'amertume des peroles que vient de prononcer À gamemnon ; Agamemnon répond comme si Ulysse avait eu en vue l’amertume de ses propres remontrances, 4866.’EvOaë’ Tlouet, j'en arriverai là, c'est-à-dire, je mourrai, .Dindorf : Kai' \textgreek{αὑτὸς ἴξομαι πρὸς τὸ θάπτειν αὐτόν}.}}, the name of a person \textit{\textsf{Agamemnon}} has been correctly identified, however, \textit{\textsf{À gamemnon}} has not been recognized. Not surprisingly, when the entity is highly impacted by the OCR process, such as \textit{\textsf{Becthoven}} instead of \textit{\textsf{Beethoven}} in \textit{\textsf{XIe Quatuor (op. 95) Becthoven}}, ChatGPT fails to detect such entities.

Results also showed  that ChatGPT can be affected by spelling variations over years, with 78\% of named entities from $1958$ to $2018$ having been correctly recognized. However, this rate drops to half with entities from $1798$ to $1948$ in the \texttt{hipe-2020} dataset. 







Finally, due to an abundant amount of noise, such as in \textit{\textsf{m \_'¦'— "..'i -i' i' —'i m Nota. Les Avis à insérer dans cette Feuille, qui ne seront pas remis au Bureau le Mardi matin, à neuf-heures au plus tard, seront renvoyés irrévocablement à l'ordinaire prochain \_. ARTICLES OFFICIELS. k. Le public est informé, qu'au bénéfice d'un gracieux arrêt de la Seigneurie, "et d'une sentence de direction de l'honorable Justice de la Chaux-de-Fonds, le tuteur et les parens des enfans' de Henri Jean-Maire, se présenteront en cour de Justice de la die Chatùx-de-Fonds, le Mardi \_s \_? }}, ChatGPT ``surrenders'' with this statement: \textit{\textsf{The text you provided is not in a coherent language, and it is difficult to understand what it is trying to communicate. The text contains mostly punctuation and special characters with no recognizable entities. }}

\subsection{Code-switching}


Code-switching refers to the phenomenon of alternating between two or more languages within the same sentence, phrase, or single word. The monolingual bias in multilingualism is a type of bias that can occur in language models \cite{talat2022you,choudhury2021linguistically}. Not only, but these LMs are also hardly generalizable to different code-switched languages \cite{aguilar2020lince} and pre-trained multilingual models do not necessarily guarantee high-quality representations on code-switching \cite{winata2021multilingual}. 

\texttt{ajmc} presents code-switching between French and Ancient Greek. GPT-3.5 was trained in more than 100 languages, being English over-represented with 93\% word count, 1.82\% for French and 0.032\% for Modern Greek, while Ancient Greek is unrepresented \cite{brown2020language}. ChatGPT was trained with further datasets for fine-tuning, human feedback and prompting \cite{ouyang2022training} but similar language distribution is to be expected. 
Thus, for an example such as \textit{\textsf{À la marge d’an exemplaire de Sophocle, on lit la traduction suivante de ces deux vers, due à notre ÆRacine : « O mon fils, sois un jour plas beareux que ton pèren; Da reste avec honneur tu peux lui ressembler». }} ChatGPT ``confidently'' responds with \textit{\textsf{I'm sorry, but I'm unable to understand the text you have provided. The text appears to be a mixture of ancient Greek and French, with some references to ancient literature and annotations, which makes it difficult to extract meaningful information. Additionally, the text is not a historical text, but rather a literary text, which also makes it difficult to extract historical entities.}}

\section{Conclusions \& Perspectives}
On this note, we conclude that ChatGPT encounters several difficulties in recognizing entities in historical documents, that range from the consistency of entity annotation guidelines, entity complexity, multilingualism and code-switching, to the specificity of prompting. Moreover, while an unprecedented amount of historical documents is available in digital format, little is freely available with many historical archives that remain inaccessible to the public (and consequently, on the Internet). For example, primary sources (documents or other items that provide first-hand, eyewitness accounts of events) such as newspapers and magazine articles (as in the case of the majority of the datasets from this study) are available both online and in the library, but nevertheless, they are generally watermarked or behind a paywall. Consequently, ChatGPT is ``unaware'' (for now) of such knowledge, which contributes to the degree of confusion of the model with regard to historical documents. Even if these resources become accessible and LLMs-based systems improve their ``understanding'' capacities in historical documents, their implementation in digital libraries must be taken with caution to prevent biased and out-of-domain responses.

\section*{Ethics Statement}
While it can generate plausible-sounding text, the content generated by ChatGPT is not necessarily true. Nevertheless, we consider that it does not concern the task of named entity recognition, as we are not adding any further ethical consideration other than those posed by ChatGPT. We are aware of the intentional stance \cite{dennett2009intentional} of terms such as ``confidently'', ``surrenders'', ``understanding'' and ``unaware'' when applied to a conversational agent, however, in this paper, we adopt them to emphasize ChatGPT capacity to interact with a user.

\section*{Acknowledgements}

This work has been supported by the  ANNA (2019-1R40226) and TERMITRAD (2020-2019-8510010) projects funded by the Nouvelle-Aquitaine Region, France.


%

%


\bibliographystyle{ACM-Reference-Format}
\bibliography{sample-base}


\end{document}